 \documentclass[preprint]{aastex}

\slugcomment{NAOJ-Th-Ap 2000,No.33; ApJ, in press}

\shorttitle{Yutaka Fujita}
\shortauthors{Ram-Pressure Stripping}

\begin{document}

\title{Ram-Pressure Stripping of Galaxies in High-Redshift
Clusters and Influence of ICM Heating}

\author{Yutaka Fujita}
\affil{National
Astronomical Observatory, Osawa 2-21-1, Mitaka, Tokyo 181-8588, Japan}
\email{yfujita@th.nao.ac.jp}

\begin{abstract}
 We have investigated the ram-pressure stripping of disk galaxies in
 clusters at various redshifts and in cluster progenitors; the clusters
 grow up on a hierarchical clustering scenario. We consider a radially
 infalling galaxy whose initial position and velocity are given by a
 spherical collapse model of structure formation. Moreover, since
 observations show that the intracluster medium (ICM) of nearby clusters
 is non-gravitationally heated, we study the effect of the
 non-gravitational heating on the ram-pressure stripping. For a given
 redshift, we find that ram-pressure stripping has more influence on
 galaxies in more massive clusters. On the other hand, for a given mass,
 it has more influence on galaxies in the clusters at higher
 redshifts. In particular, we predict that in rich clusters at $z\gtrsim
 1$, most of the galaxies are affected by the ram-pressure
 stripping. While the non-gravitational heating significantly reduces
 the influence of ram-pressure stripping on galaxies in clusters with
 mass smaller than $1-5\times 10^{13}\;\rm M_{\sun}$, it does not affect
 the influence in richer clusters. If the ICM is heated
 non-gravitationally at $z>>1$, ram-pressure stripping does not occur at
 $z\sim 1-2$ in the progenitors of clusters observed at $0\lesssim z
 \lesssim 0.5$, because the heat makes the ICM fraction of the cluster
 progenitors small. On the other hand, if the ICM is heated
 non-gravitationally at $z\sim 0$ for the first time, the ram-pressure
 stripping occurs even at $z\sim 3$. We compare the results with the
 observations of galaxies in clusters at various redshifts.
\end{abstract}


\section{Introduction}
\label{sec:intro}

By comparing observations of galaxies in clusters at $z\gtrsim 0.2$ with
those in clusters at $z\sim 0$, we infer that some environmental effects
in clusters have influences on the evolution of the galaxies.
\citet{but78,but84} found that clusters at $z\gtrsim 0.2$ have a high
fraction of blue galaxies in comparison with nearby clusters, and
subsequent works have confirmed this trend
\citep[e.g.][]{lub96,rak95,ell00}. Recent observations with the {\em
Hubble Space Telescope} (HST) revealed details of the blue
galaxies. \citet{dre94} and \citet{cou98} found that most of the blue
galaxies are normal spirals with active star formation. On the other
hand, observations have shown that the fraction of SO galaxies decreases
rapidly with redshift in contrast with the normal spirals
\citep{dre97,sma98,cou98}. These suggest that the blue normal spirals
observed in high redshift clusters evolve into the non-blue SO galaxies
observed in nearby clusters. In fact, observations show that in distant
clusters there are galaxies in post-starburst
phase.\citep{van98,pog99}. These galaxies may be the ones for which star
formation activity is dying down.

Several mechanisms are proposed that can lead to the color and
morphological transformations between galaxy classes in clusters, such
as galaxy mergers \citep{bek98}, tides by the cluster potential
\citep{byr90,fuj98}, and tidal interactions between galaxies
\citep{moo96}. One of the strongest candidates is the ram-pressure
stripping proposed by \citet{gun72}. If a cluster galaxy moves through
hot intracluster medium (ICM), it feels ram-pressure from the ICM, which
is given by $\rho_{\rm ICM}v_{\rm rel}^2$, where $\rho_{\rm ICM}$ is the
ICM density and $v_{\rm rel}$ is the relative velocity between the
galaxy and the ICM. If the ram-pressure becomes large enough, the
interstellar medium (ISM) of the galaxy cannot be hold by the gravity of
the galaxy and is swept away. Numerical simulations demonstrate that in
a cluster environment, the stripping is likely to occur
\citep{tak84,gae87,por93,aba99,mor00}. In particular, high-resolution
three dimensional numerical simulations show that the ram-pressure
stripping is so effective that it removes 100\% of the atomic hydrogen
content of luminous galaxies within $10^8$ yr \citep{qui00}. On the
other hand, \citet{fuj99a} investigated the influence of ram-pressure on
the star formation activity of a disk galaxy. They found that just
before the atomic hydrogen content is stripped, the star formation rate
increases at most a factor of 2, but rapidly decreases on a timescale of
$10^8$ yr after the stripping. After the star formation activity, which
mainly occurred in the disk, ceases, the galaxy looks like a S0 galaxy
in both color and morphology \citep{fuj99a}. HI deficient galaxies and
galaxies with no strong emission-lines seen in cluster cores support the
theoretical predictions \citep[e.g.][]{gio85,dal00}. Although
ram-pressure stripping alone does not explain the detailed morphological
features of S0 galaxies, such as their large bulge to disk ratios or
their conspicuous thick disks \citep{bur79a,bur79b}, it may be a
principal mechanism of the transformation of spirals with active star
formation into S0 galaxies with inactive star formation.

However, most of the previous studies dealt with the ram-pressure
stripping of a model galaxy in a given cluster with arbitrary initial
conditions. Moreover, they did not take the evolution of cluster
structure into account; as will be seen in \S\ref{sec:model}, the
structure of high-redshift clusters is different from that of nearby
clusters even for the same mass. Since it affects the ICM density, the
velocity of the galaxies, and the efficiency of ram-pressure stripping,
it must be considered when we compare the theoretical models with
observations of high-redshift clusters. In this paper, we investigate
ram-pressure stripping in clusters at various redshifts, which grow 
according to a hierarchical clustering scenario; the initial position
and velocity of galaxies are given by a spherical collapse model of
cluster formation. Moreover, since a cluster breaks into smaller
progenitors as time goes backwards, galaxies in the cluster might have
been affected by ram-pressure stripping when they were in the
progenitors before the present-day cluster formed. Thus, we also
consider the ram-pressure stripping of galaxies in these progenitors.

Since ram-pressure is proportional to the density of ICM, the ICM
distribution of a cluster may be related to the evolution of the cluster
galaxies feeling ram-pressure. X-ray observations of nearby clusters
show that their ICM distributions are generally different from their
dark matter distributions. In particular, for low temperature clusters,
the distributions of ICM are significantly flatter than those of dark
matter and the ICM fraction in their central regions is small
\citep{moh99,ett99,arn99}. A possible explanation of the ICM
distributions is that the ICM has been heated non-gravitationally. In
fact, \citet{pon99} indicated that the entropy of the ICM in the central
regions of low-temperature or less massive clusters is higher than can
be explained by gravitational collapse alone, although it is not
understood what heats the ICM (e.g. supernovae or AGN) and where the ICM
is heated, that is, outside or inside clusters. Heating other than the
gravity of a cluster makes the ICM distribution flatter and different
from the dark matter distribution. Thus, we expect that the position
where a galaxy suffers from ram-pressure stripping depends on whether
the ICM of the cluster (or the gas accreted by the cluster later on) has
been heated non-gravitationally or not. In particular, we expect that
the position where ram-pressure stripping occurs is more sensitive to
the non-gravitational heating in the past. This is because a cluster
breaks into progenitors or less massive clusters as time goes backwards,
and because the heat required to explain the observations should have
more influence on the ICM distributions of the less massive clusters
\citep{cav98,bal99a,loe00,fuj00}. Therefore, ram-pressure stripping in
the progenitors may tell us when the ICM of present-day clusters was
heated non-gravitationally; it will be a clue to the heating sources.

Our paper is organized as follows. In \S\ref{sec:model} we describe our
models of the dark matter distribution and the ICM distribution within
clusters, the ram-pressure stripping of a radially infalling galaxy, and
the evolution of cluster progenitors. In \S\ref{sec:result} we give the
results of our calculations. We compare them with observations in
\S\ref{sec:disc}. Conclusions are given in \S\ref{sec:conc}.

\section{Models}
\label{sec:model}

\subsection{Distributions of Gravitational Matter and ICM}
\label{sec:dist}

The virial radius of a cluster with virial mass $M_{\rm vir}$ is defined
as
\begin{equation}
 \label{eq:r_vir}
r_{\rm vir}=\left(\frac{3\: M_{\rm vir}}
{4\pi \Delta_c(z) \rho_{\rm crit}(z)}\right)^{1/3}\:,
\end{equation}
where $\rho_{\rm crit}(z)$ is the critical density of the universe and
$\Delta_c(z)$ is the ratio of the average density of the cluster to the
critical density at redshift $z$. The former is given by
\begin{equation}
\label{eq:rho_crit}
 \rho_{\rm crit}(z)
=\frac{\rho_{\rm crit,0}\Omega_0 (1+z)^3}{\Omega(z)}\:,
\end{equation} 
where $\rho_{\rm crit,0}$ is the critical density at $z=0$, and 
$\Omega(z)$ is the cosmological density parameter. The latter
is given by
\begin{equation}
\label{eq:Dc}
  \Delta_c(z)=18\:\pi^2 
\end{equation}
for the Einstein-de Sitter Universe and
\begin{equation}
\label{eq:Dc_lam}
  \Delta_c(z)=18\:\pi^2+82 x-39 x^2\:, 
\end{equation}
for the flat universe with non-zero cosmological constant
\citep{bry98}. In equation (\ref{eq:Dc_lam}), the parameter $x$ is given
by $x=\Omega(z)-1$.

We assume that a cluster is spherically symmetric and the density
distribution of gravitational matter is
\begin{equation}
 \label{eq:rho_m}
\rho_{\rm m}(r)=\rho_{\rm mv}(r/r_{\rm vir})^{-\alpha}\:,
\end{equation}
where $\rho_{\rm mv}$ and $\alpha$ are constants, and $r$ is the
distance from the cluster center. The normalization, $\rho_{\rm mv}$, is
given by
\begin{equation}
\rho_{\rm mv}=\frac{3-\alpha}{3}\Delta_c\rho_{\rm crit}\:. 
\end{equation}
We choose $\alpha=2.4$, because the slope is consistent with
observations \citep{hor99}. Moreover, the results of numerical
simulations show that the mass distribution in the outer region of
clusters is approximately given by equation (\ref{eq:rho_m}) with
$\alpha \sim 2.4$ \citep{nav96,nav97}.

We consider two ICM mass distributions. One follows equation
(\ref{eq:rho_m}) except for the normalization and the core structure;
\begin{equation}
\label{eq:ICM_G}
 \rho_{\rm ICM}(r)=\rho_{\rm ICM, vir}
\frac{[1+(r/r_{\rm c})^2]^{-\alpha/2}}
{[1+(r_{\rm vir}/r_{\rm c})^2]^{-\alpha/2}}\:.
\end{equation}
The ICM mass within the virial radius of a cluster is
\begin{equation}
\label{eq:M_ICM}
 M_{\rm ICM}=\int_{0}^{r_{\rm vir}}4 \pi r^2 \rho_{\rm ICM}(r)dr\:.
\end{equation}
The normalization $\rho_{\rm ICM, vir}$ is determined by the relation
$f_{\rm b}=M_{\rm ICM}/M_{\rm vir}$.  where $f_{\rm b}$ is the gas or
baryon fraction of the universe. This distribution corresponds to the
case where the ICM is in pressure equilibrium with the gravity of the
cluster and is not heated by anything other than the gravity. We
introduce the core structure to avoid the divergence of gas density at
$r=0$ and use $r_{\rm c}/r_{\rm vir}=0.1$. We call this distribution the
'non-heated ICM distribution'. We use $f_{\rm b}=0.25 (h/0.5)^{-3/2}$,
where the present value of the Hubble constant is written as
$H_0=100\:h\rm\: km\:s^{-1}\: Mpc^{-1}$. The value of $f_b$ is the
observed ICM mass fraction of high-temperature clusters
\citep{moh99,ett99,arn99}, for which the effect of non-gravitational
heating is expected to be small.

However, as mentioned in \S\ref{sec:intro}, observations show that ICM
is additionally heated non-gravitationally at least for nearby
clusters. Thus, we also model the distribution of the heated ICM using
the observed parameters of nearby clusters as follows. In this paper, we
assume that the ICM had been heated before accreted by
clusters. However, the distribution will qualitatively be the same even
if the ICM is heated after accreted by clusters \citep[see][]{loe00}.

Following \citet{bal99a}, we define the adiabat $K_0=P/\rho_{\rm
g}^{\gamma}$, where $P$ is the gas pressure, $\rho_{\rm g}$ is its
density, and $\gamma$ is a constant. If ICM had already been heated
before accreted by a cluster, the entropy prevents the gas from
collapsing into the cluster with dark matter. In this case, the ICM
fraction of the cluster is given by
\begin{equation}
\label{eq:f_ICM}
f_{\rm ICM}=\min
\left[0.040\left(\frac{M_{\rm vir}}{10^{14}\:\rm M_{\sun}}\right)
\left(\frac{K_0}{K_{34}}\right)^{-3/2}
\left(\frac{t(z)}{10^9\rm\: yr}\right),
\:f_{\rm b}\right]\:,
\end{equation}
where $K_{34}=10^{34}\rm\: erg\:g^{-5/3}\:cm^{2}$ \citep{bal99a}.

The virial temperature of a cluster is given by
\begin{equation}
 \frac{k_{\rm B}T_{\rm vir}}{\mu m_{\rm H}}
=\frac{1}{2}\frac{GM_{\rm vir}}{r_{\rm vir}}\:,
\end{equation}
where $k_{\rm B}$ is the Boltzmann constant, $\mu (=0.61)$ is the mean
molecular weight, $m_{\rm H}$ is the hydrogen mass, and $G$ is the
gravitational constant. When the virial temperature of a cluster is much
larger than that of the gas accreted by the cluster, a shock forms near
the virial radius of the cluster \citep{tak98,cav98}. The temperature of
the postshock gas ($T_2$) is related to that of the preshock gas ($T_1$)
and is approximately given by
\begin{equation}
 T_2 = T_{\rm vir}+\frac{3}{2} T_1
\end{equation}
\citep{cav98}. Since the gas temperature does not change very much for
$r<r_{\rm vir}$ (\citealp{tak98}; see also \citealp{irw00}), the ICM
temperature of the cluster is given by $T_{\rm ICM}=T_2$. Since we
assume that the density profile of gravitational matter is given by
equation (\ref{eq:rho_m}) with $\alpha=2.4$, the density profile of ICM
is given by
\begin{equation}
\label{eq:ICM_H}
 \rho_{\rm ICM}(r)=\rho_{\rm ICM, vir}
\frac{[1+(r/r_{\rm c})^2]^{-3\beta/2}}
{[1+(r_{\rm vir}/r_{\rm c})^2]^{-3\beta/2}}\:, 
\end{equation}
where $\beta=(2.4/3)T_{\rm vir}/T_{\rm ICM}$
\citep[see][]{bah94}. Observations show that $T_1\sim 0.5-1$~keV
although it depends on the distribution of the gravitational matter in a
cluster \citep{cav98,fuj00}. We choose $3T_{1}/2=0.8$~keV, hereafter.
The normalization $\rho_{\rm ICM, vir}$ is determined by the relation
$f_{\rm ICM}=M_{\rm ICM}/M_{\rm vir}$.

When $T_{\rm vir}\lesssim (3/2)T_1$, a shock does not form but the gas
accreted by a cluster adiabatically falls into the cluster. The ICM
profile for $r<r_{\rm vir}$ is obtained by solving the equation of
hydrostatic equilibrium,
\begin{equation}
\label{eq:stat}
 \frac{dP}{dr}=-\frac{GM(r)}{r^2}\rho_{\rm ICM}\:,
\end{equation}
where $M(r)$ is the mass of the cluster within radius $r$. Generally,
equation (\ref{eq:stat}) does not have analytical solutions for the
matter distribution we adopted (equation~[\ref{eq:rho_m}] with
$\alpha=2.4$). Thus, we use the solution for the isothermal distribution
($\rho_{\rm m}[r]=\rho_{\rm mv,iso} [r/r_{\rm vir}]^{-2}$) as an
approximation; assuming $\gamma=5/3$, it is given by
\begin{equation}
\label{eq:rho_ad}
 \rho_{\rm ICM}(r)=\rho_{\rm ICM, vir}
\left[1+\frac{3}{A}
\ln\left(\frac{r_{\rm vir}}{r}\right)\right]^{3/2}\:,
\end{equation}
where 
\begin{equation}
 A=\frac{15 K_0 \rho_{\rm ICM, vir}^{2/3}}
{8\pi G \rho_{\rm mv,iso}r_{\rm vir}^2}\:
\end{equation}
\citep{bal99a}. The parameter $\rho_{\rm ICM, vir}$ is determined by the
relation $f_{\rm ICM}=M_{\rm ICM}/M_{\rm vir}$. In \S\ref{sec:result},
we use the profile (\ref{eq:ICM_H}) for $T_{\rm
vir}>3T_{1}/2\;(=0.8\rm\; keV)$ and the profile (\ref{eq:rho_ad}) for
$T_{\rm vir}<3T_{1}/2$. We refer to this ICM distribution as `the heated
ICM distribution'. From observations, we use the value of $K_0=0.37
K_{34}$, which is assumed to be independent of cluster mass
\citep{bal99a}.

\subsection{A Radially Infalling Galaxy}
\label{sec:gal}

We consider a radially infalling disk galaxy accreted by a cluster with
dark matter. The initial velocity of the model galaxy, $v_i$, is given
at $r=r_{\rm vir}$, and it is
\begin{equation}
 \frac{v_i^2}{2}=\frac{G M_{\rm vir}}{r_{\rm vir}}
-\frac{G M_{\rm vir}}{r_{\rm ta}}\:,
\end{equation}
where $r_{\rm ta}$ is the turnaround radius of the cluster. Assuming
that $r_{\rm ta}=2 r_{\rm vir}$ on the basis of the virial theorem, the
initial velocity is
\begin{equation}
 v_i = \sqrt{\frac{G M_{\rm vir}}{r_{\rm vir}}}\:.
\end{equation}
The virial radius $r_{\rm vir}$ is given by equation (\ref{eq:r_vir}).

The velocity of the model galaxy is obtained by solving the equation of
motion;
\begin{equation}
 \label{eq:motion}
\frac{d^2 r}{dr^2}=-\frac{G M(r)}{r^2}\:.
\end{equation}
As the velocity of the galaxy increases, the ram-pressure from ICM also
increases. The condition of ram-pressure stripping is
\begin{eqnarray}
  \label{eq:strip}
  \rho_{\rm ICM}v_{\rm rel}^2 
 & >& 2\pi G \Sigma_{\star} \Sigma_{\rm HI} \nonumber\\
 & =& v_{\rm rot}^2 R^{-1} 
  \Sigma_{\rm HI} \label{eq:grav2} \nonumber\\
 & =& 2.1\times 10^{-11}{\rm dyn\: cm^{-2}}
               \left(\frac{v_{\rm rot}}{220\rm\; km\: s^{-1}}\right)^2
               \nonumber\\
 &  &   \times \left(\frac{R}{10\rm\; kpc}\right)^{-1}
               \left(\frac{\Sigma_{\rm HI}}
               {8\times 10^{20} 
                   m_{\rm H}\;\rm cm^ {-2}}\right) \label{eq:grav3}\:, 
\end{eqnarray}
where $\Sigma_{\star}$ is the gravitational surface mass density,
$\Sigma_{\rm HI}$ is the surface density of the HI gas, $v_{\rm rot}$ is
the rotation velocity, and $R$ is the characteristic radius of the
galaxy \citep{fuj99a, gun72}. \citet{aba99} have numerically confirmed
that this analytic relation provides a good approximation. We define the
cluster radius at which the condition (\ref{eq:strip}) is satisfied for
the first time as the stripping radius, $r_{\rm st}$. Since we assume
that the ICM is nearly in pressure equilibrium for $r<r_{\rm vir}$, the
relative velocity $v_{\rm rel}$ is equivalent to the velocity of the
galaxy relative to the cluster for $r<r_{\rm vir}$.

Since the mass distribution of a cluster within the virial radius does
not change rapidly, 
the time that a galaxy takes from $r_{\rm vir}$ to
$r_{\rm st}$ is approximately given by
\begin{equation}
\Delta t= \int_{r_{\rm st}}^{r_{\rm vir}}
\frac{dr}{\sqrt{v_i^2-2\Delta\phi(r)}}\:,
\end{equation}
where
\begin{equation}
\Delta\phi(r)
=\frac{4\pi G \rho_{\rm mv}r_{\rm vir}^{\alpha}}{(2-\alpha)(3-\alpha)}
(r^{2-\alpha}-r_{\rm vir}^{2-\alpha})\:.
\end{equation} 
Thus, when the ISM of the galaxy that was located at $r=r_{\rm vir}(t')$
at $t=t'$ is stripped at $r=r_{\rm st}$ at $t=t'+\Delta t$, the virial
radius of the cluster becomes larger than $r_{\rm vir}(t')$.

\subsection{The Growth of Clusters}
\label{sec:cluster}

We also investigate ram-pressure stripping in progenitors of a
cluster. In this subsection, we construct a model of the growth of
clusters. 

The conditional probability that a particle which resides in a object
(`halo') of mass $M_2$ at time $t_2$ is contained in a smaller halo of
mass $M_1\sim M_1+d M_1$ at time $t_1$ ($t_1<t_2$) is
\begin{equation}
 \label{eq:prob}
P_{1}(M_1,t_1|M_2,t_2)d M_1
=\frac{1}{\sqrt{2\pi}}
\frac{\delta_{c1}-\delta_{c2}}{(\sigma_{1}^2-\sigma_{2}^2)^{3/2}}
\left|\frac{d\sigma_1^2}{d M_1}\right|
\exp\left[\frac{(\delta_{c1}-\delta_{c2})^2}
{2(\sigma_{1}^2-\sigma_{2}^2)}\right]d M_1 \;,
\end{equation}
where $\delta_{ci}$ is the critical density threshold for a spherical
perturbation to collapse by the time $t_i$, and $\sigma_i [\equiv
\sigma(M_i)]$ is the rms density fluctuation smoothed over a region of
mass $M_i$ for $i=1$ and 2 \citep{bon91,bow91,lac93}.  In this paper, we
use an approximative formula of $\delta_{ci}$ \citep{nak97} and an
fitting formula of $\sigma(M)$ for the CDM fluctuation spectrum
\citep{kit98}.

We define the typical mass of halos at $t$ that become part of a larger
halo of mass $M_0$ at later time $t_0 (>t)$ as
\begin{equation}
 \label{eq:m_ave}
 \bar{M}(t|M_0,t_0)
=\frac{\int_{M_{\rm min}}^{M_0} M P_{1}(M,t|M_0,t_0)d M}
{\int_{M_{\rm min}}^{M_0} P_{1}(M,t|M_0,t_0)d M} \:,
\end{equation}
where $M_{\rm min}$ is the lower cutoff mass. We choose $M_{\rm
min}=10^8\rm\: M_{\sun}$, which corresponds to the mass of dwarf
galaxies.

In \S\ref{sec:prog}, we investigate a cluster progenitor whose
virial mass is given by
\begin{equation}
 \label{eq:m_vir}
M_{\rm vir}(t|M_0,t_0)=\bar{M}(t|M_0,t_0) \:.
\end{equation}
In that subsection, we will often represent $M_{\rm vir}(t|M_0,t_0)$
with $M_{\rm vir}(t)$ or $M_{\rm vir}$ unless it is misunderstood, and
we will often abbreviate a cluster progenitor to a `cluster'.

\section{Results}
\label{sec:result}

The parameters for a model galaxy are $v_{\rm rot}=220\:\rm km\:
s^{-1}$, $R=10$ kpc, and $\Sigma_{\rm HI}=8\times 10^{20}m_{\rm H}\rm\:
cm^{-2}$. Although the values are those for our Galaxy \citep{spi78,
bin87}, the condition of ram-pressure stripping (relation
[\ref{eq:strip}]) is not much different even in the case of smaller
galaxies. Taking M33 for instance, the relation (\ref{eq:strip}) turns
to be $\rho_{\rm ICM}v_{\rm rel}^2 > 2.2\times 10^{-11}{\rm dyn\:
cm^{-2}}$, because the values are $v_{\rm rot}=105\:\rm km\: s^{-1}$,
$R=5$ kpc, and $\Sigma_{\rm HI}=18\times 10^{20}m_{\rm H}\rm\: cm^{-2}$
\citep{new80,gar97,sof99}.

As cosmological models, we consider a standard cold dark mater model
(SCDM model) and a cold dark matter model with non-zero cosmological
constant ($\Lambda$CDM model). The cosmological parameters are $h=0.5$,
$\Omega_0=1$, $\lambda_0\equiv \Lambda/(3 H_0^2)=0$, and $\sigma_8=0.63$
for the SCDM model, and $h=0.75$, $\Omega_0=0.25$, $\lambda_0=0.75$, and
$\sigma_8=1.0$ for the $\Lambda$CDM model. As will be seen, the results
do not much depend on the cosmological models.

\subsection{The Relation Between Mass and Stripping Radius}
\label{sec:m_r}

\begin{figure}
\figurenum{1}
\epsscale{0.80}
\plotone{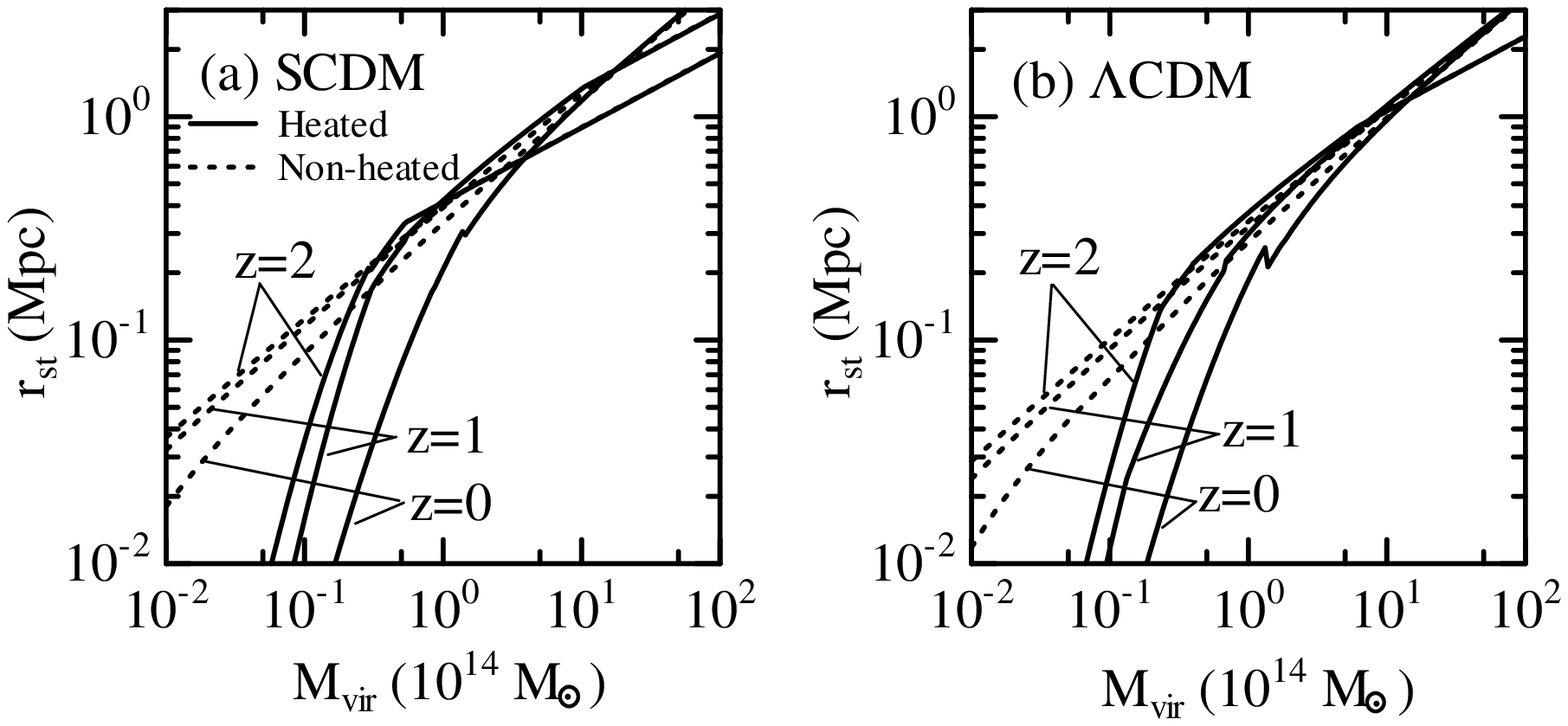}
\caption{The relations between the virial mass and the
stripping radius of clusters at $z=0$, 1, and 2. (a) SCDM, (b)
$\Lambda$CDM. The solid and dotted lines indicate the heated and the
non-heated ICM models, respectively.  \label{fig:m_rst}}
\end{figure}

\begin{figure}
\figurenum{2}
\epsscale{0.80}
\plotone{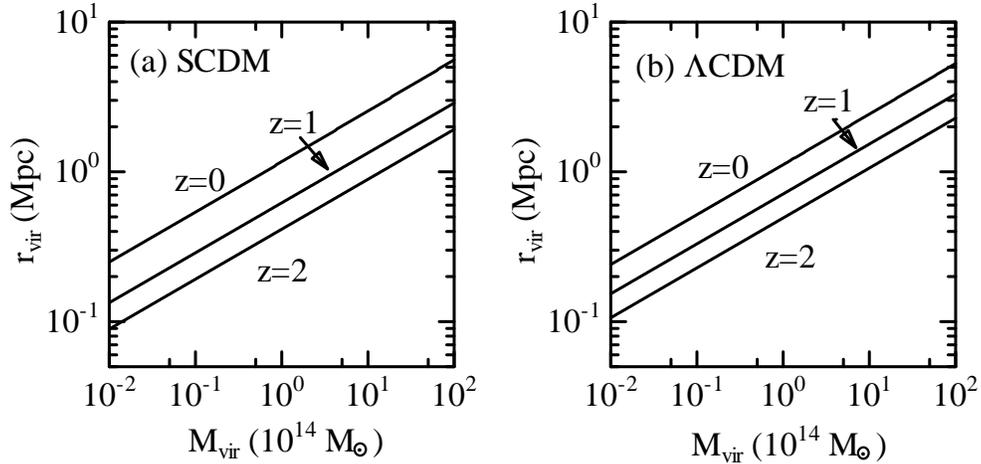}
\caption{The relations between the virial mass and the virial
radius of clusters at $z=0$, 1, and 2. (a) SCDM, (b)
$\Lambda$CDM. \label{fig:m_rvir}}
\end{figure}

We first discuss the results of the non-heated ICM
models. Figure~\ref{fig:m_rst} shows the relations between the stripping
radius, $r_{\rm st}$, and virial mass, $M_{\rm vir}$, of clusters at
several redshifts. For a given redshift, $r_{\rm st}$ is an increasing
function of $M_{\rm vir}$. One of the reasons is the mass-dependence of
$r_{\rm vir}$ (Figure~\ref{fig:m_rvir}). That is, massive clusters are
large. To see the effect of ram-pressure stripping, we illustrate
$r_{\rm st}/r_{\rm vir}$ in Figure~\ref{fig:m_rstvir}. For a given
redshift, $r_{\rm st}/r_{\rm vir}$ is larger for more massive
clusters. Since not all galaxies have the pure radial orbits in real
clusters, we suppose that in the real clusters with larger $r_{\rm
st}/r_{\rm vir}$, more galaxies are affected by ram-pressure
stripping. In that sense, ram-pressure stripping is more effective in
more massive clusters. The mass dependence can be explicitly shown as
follows. For a given redshift, the virial density of clusters,
$\rho_{\rm vir}(\equiv \Delta_c \rho_{\rm crit})$, does not depend on
mass (equations~[\ref{eq:rho_crit}]-[\ref{eq:Dc_lam}]). Thus, the virial
radius of a cluster is given by $r_{\rm vir}\propto (M_{\rm
vir}/\rho_{\rm vir})^{1/3}\propto M_{\rm vir}^{1/3}$. Therefore the
velocity of a galaxy infalling into the cluster has the relation,
$v_i^2\propto M_{\rm vir}/r_{\rm vir}\propto M_{\rm vir}^{2/3}$. Since
the typical ICM density of a cluster is proportional to the virial
density, the ram-pressure of the galaxy follows the relation, $\rho_{\rm
ICM}v_i^2\propto M_{\rm vir}^{2/3}$.

The virial radius, $r_{\rm vir}$, of a cluster with a given mass is
smaller at higher redshift (Figure~\ref{fig:m_rvir}), because the virial
density of the cluster $\rho_{\rm vir}$ is increases as a function of
redshift (equations~[\ref{eq:r_vir}]-[\ref{eq:Dc_lam}]). On the other
hand, $r_{\rm st}$ of a cluster with a given mass is larger at higher
redshift in the non-heated ICM models (Figure~\ref{fig:m_rst}). This is
because the typical ICM density of clusters increases as the virial
density increases. Since ram-pressure is proportional to ICM density, it
is more effective at higher redshift (Figure~\ref{fig:m_rstvir}). This
can be clarified easily in the SCDM and non-heated ICM model as
follows. The ICM density of clusters follows $\rho_{\rm ICM}\propto
\rho_{\rm vir}\propto (1+z)^3$. For a fixed mass, the virial radius is
represented by $r_{\rm vir}\propto (M_{\rm vir}/\rho_{\rm
vir})^{1/3}\propto (1+z)^{-1}$. Thus, the velocity of a galaxy is given
by $v_i^{2}\propto M_{\rm vir}/r_{\rm vir}\propto 1+z$, and the
ram-pressure is given by $\rho_{\rm ICM}v_i^2 \propto (1+z)^4$. This
explains large $r_{\rm st}/r_{\rm vir}$ of high-redshift clusters
(Figure~\ref{fig:m_rstvir}).

Next, we show the results of the heated ICM models.  In comparison with
the non-heated ICM models, $r_{\rm st}$ and $r_{\rm st}/r_{\rm vir}$ for
less massive clusters at a given redshift in the heated ICM models are
small (Figures~\ref{fig:m_rst} and \ref{fig:m_rstvir}). This is because
of the small ICM density of the model clusters. As the mass of clusters
decreases, $\beta$ in equation~(\ref{eq:ICM_H}) decreases, then the ICM
distribution changes from equation~(\ref{eq:ICM_H}) to
(\ref{eq:rho_ad}), then the ICM fraction changes from $f_{\rm
ICM}=f_{\rm b}$ to $f_{\rm ICM}<f_{\rm b}$
(equation~[\ref{eq:f_ICM}]). In Figures~\ref{fig:m_rst}
and~\ref{fig:m_rstvir}, the small jump at $M_{\rm vir}=1.3\times
10^{14}\:\rm M_{\sun}$ and $z=0$ in the $\Lambda$CDM model is due to the
shift from equation~(\ref{eq:ICM_H}) to
(\ref{eq:rho_ad}). Figure~\ref{fig:m_rstvir} shows that in the heated
ICM model, ram-pressure stripping is ineffective for galaxy clusters (or
groups) with $M_{\rm vir}\lesssim 5\times 10^{13}\rm\; M_{\sun}$ at
$z\sim 0$ and for those with $M_{\rm vir}\lesssim 10^{13}\rm\; M_{\sun}$
at $z\sim 1-2$. However, for rich clusters with $M_{\rm vir}\gtrsim
10^{14}\;\rm M_{\sun}$ observed at $z\sim 0.5-1$, the ram-pressure
stripping should be more effective than the clusters with the same mass
at $z\sim 0$ regardless of the non-gravitational heating.

\subsection{The History of Ram-Pressure Stripping in Clusters}
\label{sec:prog}

\begin{figure}
\figurenum{4}
\epsscale{0.80}
\plotone{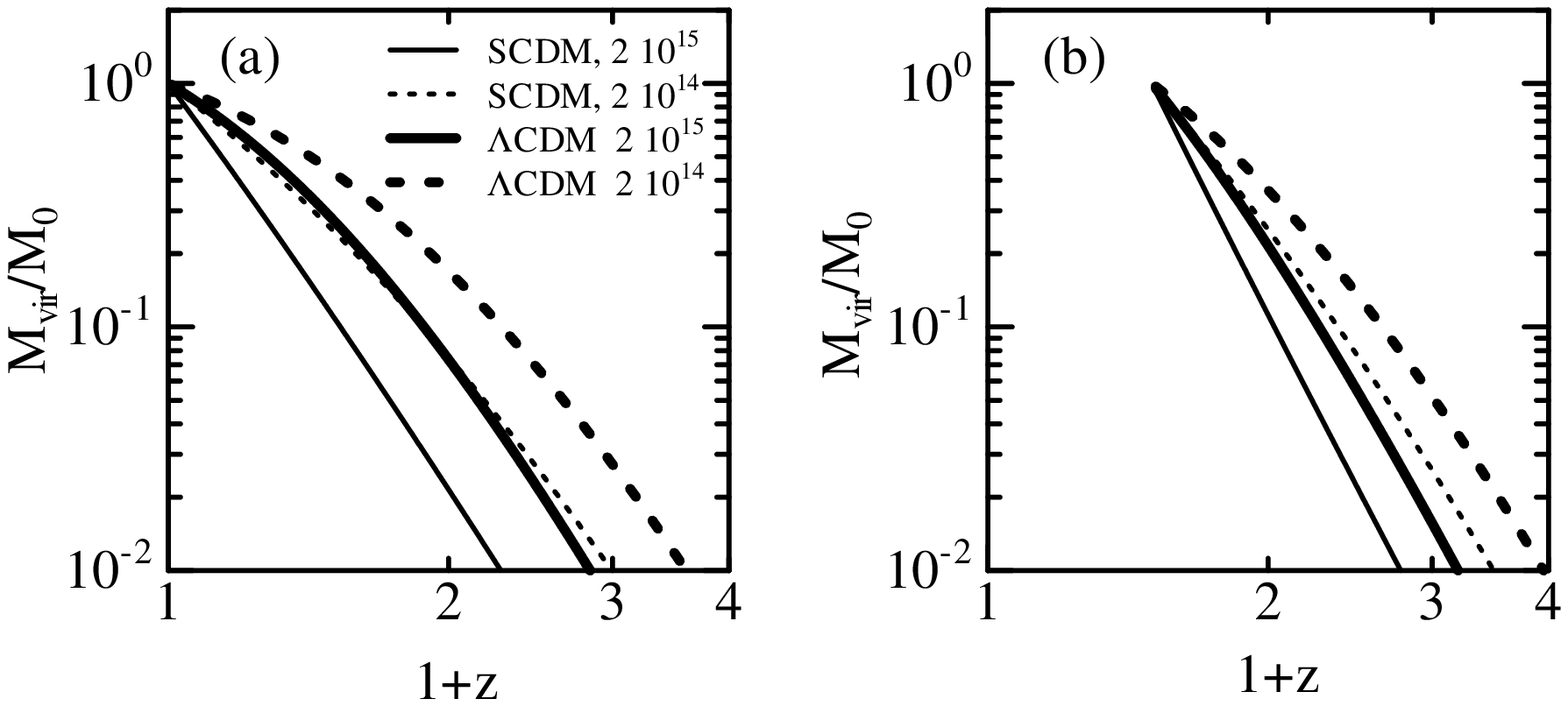}
\caption{The evolutions of the virial mass of clusters. (a)
$z_0=0$, (b) $z_0=0.5$. Four models are plotted in each figure; SCDM and
$M_0=2\times 10^{15}\:\rm M_{\sun}$ (thin solid), SCDM and $M_0=2\times
10^{14}\:\rm M_{\sun}$ (thin dotted), $\Lambda$CDM and $M_0=2\times
10^{15}\:\rm M_{\sun}$ (thick solid), and $\Lambda$CDM and $M_0=2\times
10^{14}\:\rm M_{\sun}$ (thick dotted). \label{fig:mass}}
\end{figure}

\begin{figure}
\figurenum{5}
\epsscale{0.80}
\plotone{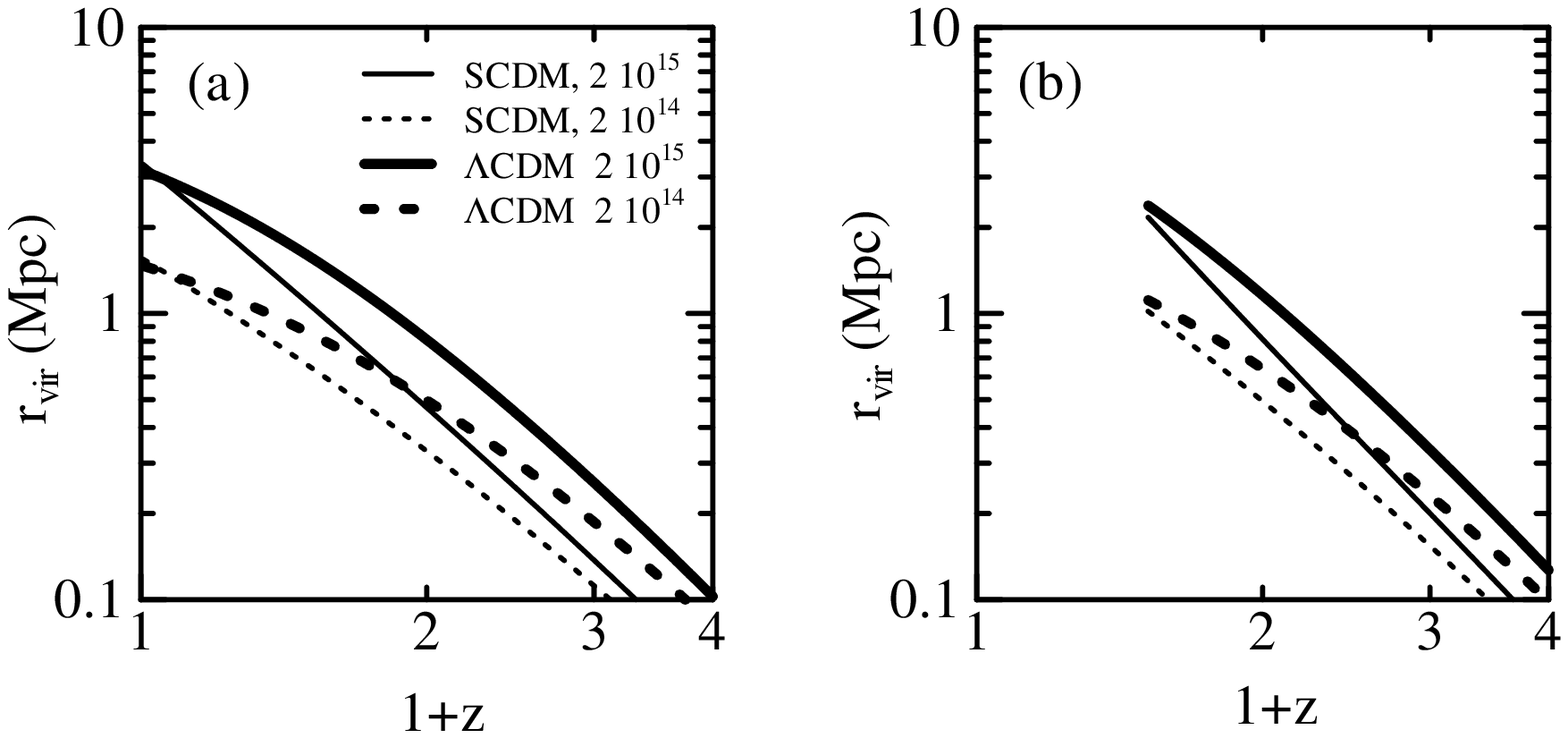}
\caption{The evolutions of the virial radius of
progenitors. (a) $z_0=0$, (b) $z_0=0.5$. Four models are plotted in each
figure; SCDM and $M_0=2\times 10^{15}\:\rm M_{\sun}$ (thin solid), SCDM
and $M_0=2\times 10^{14}\:\rm M_{\sun}$ (thin dotted), $\Lambda$CDM and
$M_0=2\times 10^{15}\:\rm M_{\sun}$ (thick solid), and $\Lambda$CDM and
$M_0=2\times 10^{14}\:\rm M_{\sun}$ (thick dotted). \label{fig:r_vir}}
\end{figure}

\begin{figure}
\figurenum{6}
\epsscale{0.80}
\plotone{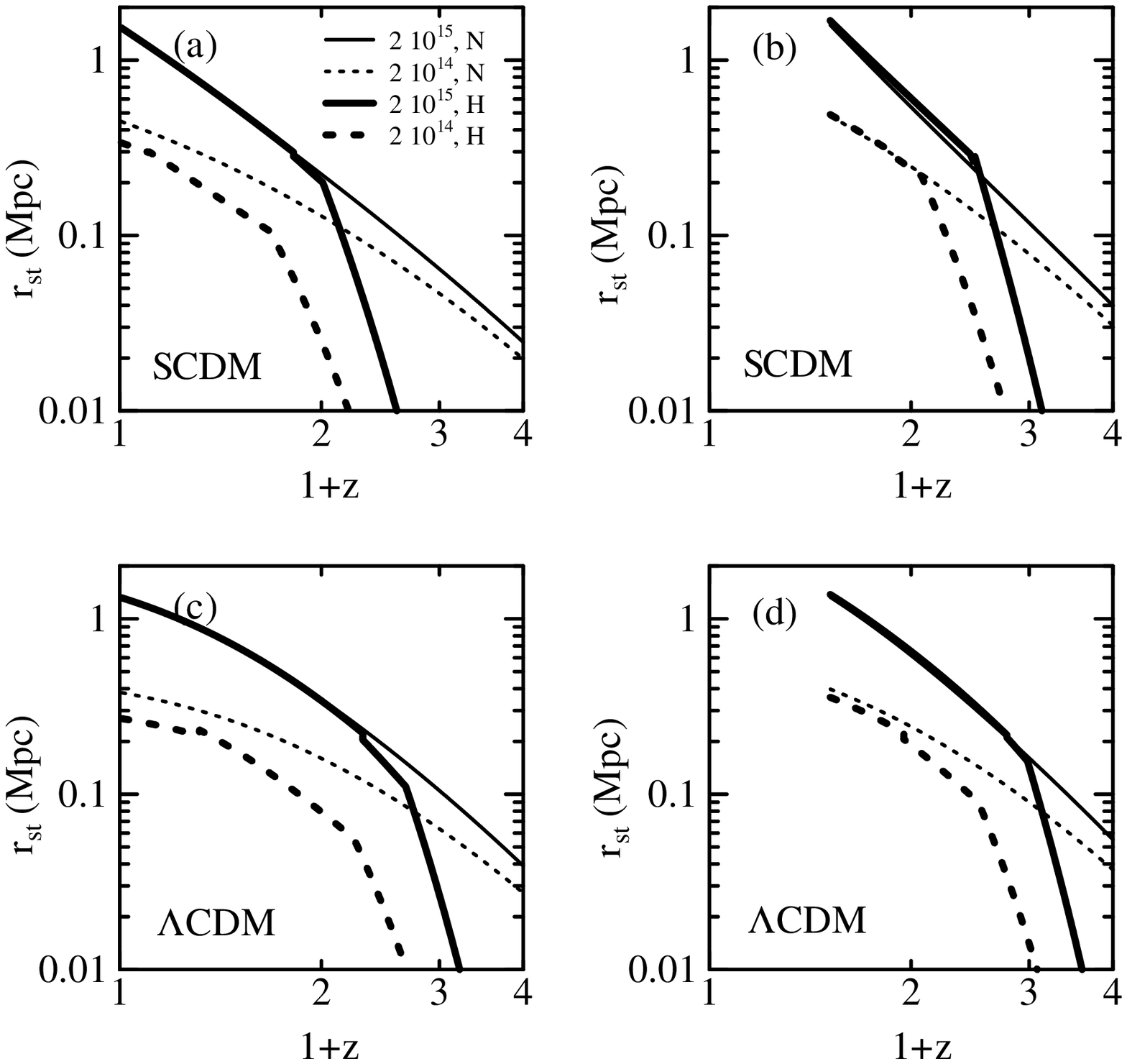}
\caption{The evolutions of the stripping radius. (a) SCDM and
$z_0=0$, (b) SCDM and $z_0=0.5$. (c) $\Lambda$CDM and $z_0=0$ (d)
$\Lambda$CDM and $z_0=0.5$. Four models are plotted in each figure;
$M_0=2\times 10^{15}\:\rm M_{\sun}$ and the non-heated ICM distribution
(thin solid), $M_0=2\times 10^{14}\:\rm M_{\sun}$ and the non-heated ICM
distribution (thin dotted), $M_0=2\times 10^{15}\:\rm M_{\sun}$ and the
heated ICM distribution (thick solid), and $M_0=2\times 10^{14}\:\rm
M_{\sun}$ and the heated ICM distribution (thick
dotted). \label{fig:r_st}}
\end{figure}

Using the model constructed in \S\ref{sec:cluster}, we investigate the
history of ram-pressure stripping of galaxies in
clusters. Figure~\ref{fig:mass} shows the evolutions of $M_{\rm
vir}(t|M_0,t_0)$ for $M_0=2\times 10^{15}\:\rm M_{\sun}$ and $2\times
10^{14}\:\rm M_{\sun}$, that is, the evolutions of clusters with typical
mass. The redshifts corresponding to $t_0$ are $z_0=0$ and 0.5. We show
the evolutions of $r_{\rm vir}$ and $r_{\rm st}$ in
Figures~\ref{fig:r_vir} and~\ref{fig:r_st}, respectively. At $z=z_0$,
the stripping radius $r_{\rm st}$ in the model of the heated ICM is not
much different from that in the model of the non-heated ICM.  In fact,
the non-gravitational heating does not affect ram-pressure stripping for
clusters with mass of $M_{\rm vir}\gtrsim 10^{14}\;\rm M_{\sun}$ at
$z\sim 0$ (Figure~\ref{fig:m_rstvir}). For clusters in that mass range,
even if the gas had been heated before accreted by clusters, the
clusters have gathered a large amount of gas and their virial
temperatures have become large enough until $z=z_0$. In our model of the
heated ICM, these respectively mean that $f_{\rm ICM}=f_{\rm b}$
(equation~[\ref{eq:f_ICM}]) and the ICM distribution is given by
equation (\ref{eq:ICM_H}). At $z=z_0$, the values of $\rho_{\rm ICM}$
are not much different between the heated ICM distribution and the
non-heated ICM distribution at $r\sim r_{\rm st}$, although the slope of
the former distribution is smaller.

\begin{figure}
\figurenum{7}
\epsscale{0.80}
\plotone{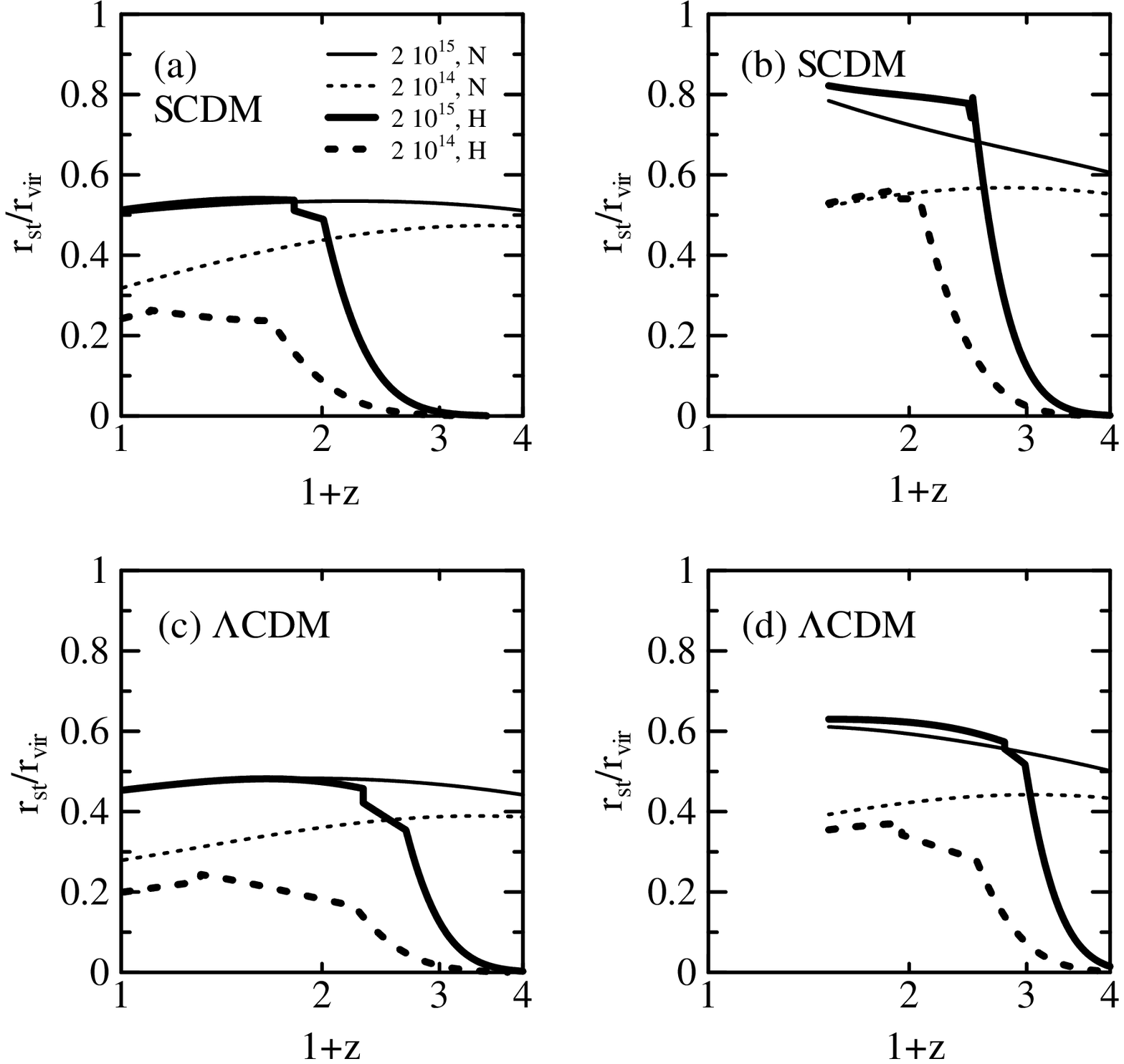}
\caption{The evolutions of the ratio of the stripping radius
to the virial radius. (a) SCDM and $z_0=0$, (b) SCDM and $z_0=0.5$. (c)
$\Lambda$CDM and $z_0=0$ (d) $\Lambda$CDM and $z_0=0.5$. Four models are
plotted in each figure; $M_0=2\times 10^{15}\:\rm M_{\sun}$ and the
non-heated ICM distribution (thin solid), $M_0=2\times 10^{14}\:\rm
M_{\sun}$ and the non-heated ICM distribution (thin dotted),
$M_0=2\times 10^{15}\:\rm M_{\sun}$ and the heated ICM distribution
(thick solid), and $M_0=2\times 10^{14}\:\rm M_{\sun}$ and the heated
ICM distribution (thick dotted). \label{fig:r_stvir}}
\end{figure}

As $z$ increases, $r_{\rm st}$ in the SCDM model decreases faster than
that in the $\Lambda$CDM model (Figure~\ref{fig:r_st}). This is because
$M_{\rm vir}$ and $r_{\rm vir}$ decrease faster in the former model
(Figures~\ref{fig:mass} and \ref{fig:r_vir}). Moreover, $r_{\rm st}$
decreases faster in the model of the heated ICM than that in the model
of the non-heated ICM. In order to see the effect of ICM heating rather
than that of the decrease of cluster size, we show the evolutions of
$r_{\rm st}/r_{\rm vir}$ in Figure~\ref{fig:r_stvir}. As can be seen,
$r_{\rm st}/r_{\rm vir}$ decreases rapidly at high redshifts in the
models of the heated ICM, while it does not change significantly in the
models of the non-heated ICM. The changes of the slope in the former
models are due to the shift in the ICM distribution from
equation~(\ref{eq:ICM_H}) to (\ref{eq:rho_ad}) and the shift in the ICM
fraction from $f_{\rm ICM}=f_{\rm b}$ to $f_{\rm ICM}<f_{\rm b}$
(equation~[\ref{eq:f_ICM}]). Taking the heated ICM model of $M_0=2\times
10^{15}\:\rm M_{\sun}$, $z_0=0$, and SCDM as an example
(Figure~\ref{fig:r_st}a and \ref{fig:r_stvir}a), the shift in the ICM
distribution occurs at $z=0.8$ and the shift in $f_{\rm ICM}$ occurs at
$z=1.0$. The rapid decrease of $r_{\rm st}/r_{\rm vir}$ in the model of
the heated ICM is chiefly attributed to the decrease of $f_{\rm
ICM}$. On the other hand, the almost constant $r_{\rm st}/r_{\rm vir}$
in the models of the non-heated ICM is explained by the fact that
although the mass of cluster progenitors and the velocity of galaxies in
them decrease with $z$ (Figure~\ref{fig:mass}), the average mass density
of progenitors, $\rho_{\rm vir}$, and thus the average ICM density of
progenitors, $f_{\rm b}\rho_{\rm vir}$, increase (equations
[\ref{eq:rho_crit}]--[\ref{eq:Dc_lam}]). We take the model of
$M_0=2\times 10^{15}\;\rm M_{\sun}$, $z_0=0$, and SCDM as an example
again.  Figures~\ref{fig:mass} and \ref{fig:r_vir} respectively show
that $M_{\rm vir}\propto (1+z)^{-6}$ and $r_{\rm vir}\propto
(1+z)^{-3}$. Thus, the velocity of the model galaxy has the redshift
dependence such as $v_i^2\propto M_{\rm vir}/r_{\rm vir}\propto
(1+z)^{-3}$. On the other hand, $\rho_{\rm ICM}\propto \rho_{\rm
vir}\propto (1+z)^3$. Therefore, the ram-pressure is almost independent
of redshift or $\rho_{\rm ICM}v_i^2\propto constant$.

In summary, Figure~\ref{fig:r_stvir} shows that if the ICM (or the gas
accreted by a cluster later on) is heated non-gravitationally at $z>>1$,
ram-pressure stripping does not occur in cluster progenitors at
$z\gtrsim 1-2$. On the other hand, if the ICM had not been heated
non-gravitationally until $z\sim 0$, ram-pressure stripping occurs even
at $z\sim 3$. The difference can be explained by the small mass of
cluster progenitors. At $z\sim 1-2$, the masses of progenitors are
$M_{\rm vir}\lesssim 10^{13}\:\rm M_{\sun}$
(Figure~\ref{fig:mass}). Figure~\ref{fig:m_rstvir} shows that the
non-gravitational heating significantly reduces the effect of
ram-pressure at this mass range at $z\sim 1-2$.

\section{Discussion}
\label{sec:disc}

In the previous sections, we have modeled the ram-pressure stripping in
clusters and their progenitors. In this section, we compare the results
with several observations. Since the direct observation of ram-pressure
stripping of galaxies, such as the observation of HI distribution in
galaxies, is difficult except for nearby clusters at present, we discuss
the morphology and color of galaxies.

\begin{figure}
\figurenum{3}
\epsscale{0.80}
\plotone{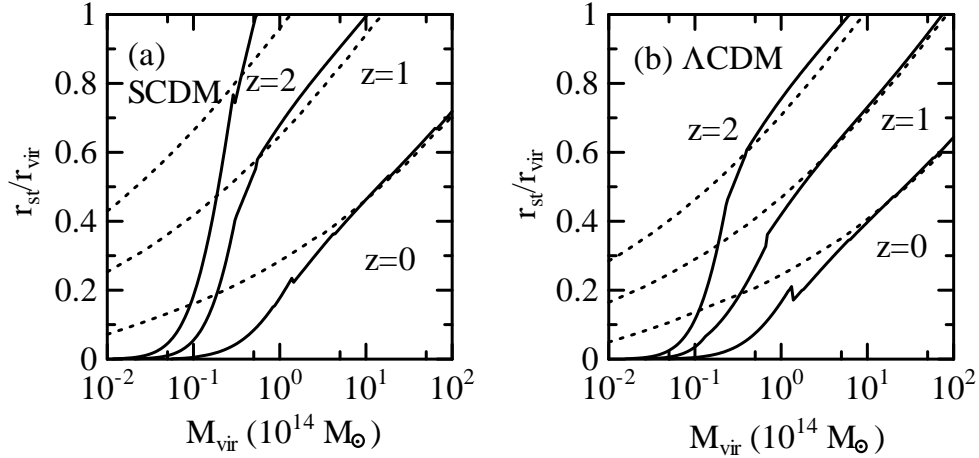}
\caption{The relations between the virial mass and the ratio
of the stripping radius to the virial radius at $z=0$, 1, and 2. (a)
SCDM, (b) $\Lambda$CDM. The solid and dotted lines indicate the heated
and the non-heated ICM models, respectively. \label{fig:m_rstvir}}
\end{figure}

In \S\ref{sec:m_r}, we show that the ram-pressure stripping in clusters
should be more effective in more massive clusters for a given redshift
(Figure~\ref{fig:m_rstvir}). \citet{edg91} investigate observational
data of nearby clusters ($z\lesssim 0.1$) and found that the spiral
fraction decreases and the S0 fraction increases with the X-ray
temperature and luminosity \citep[see also][]{bah77,mch78}. The clusters
with high X-ray temperature and luminosity are generally massive ones
\citep[e.g.][]{hor99}. Thus, the relations confirmed by \citet{edg91}
are consistent with our prediction, if the ram-pressure stripping
converts spiral galaxies into S0 galaxies. 

Recently, clusters at $z\sim 1$ are observed in detail
\citep{pos98,kaj00}. \citet{lub98} observed a cluster CL~$1604+4304$ at
$z=0.90$ and found that the fraction of S0 galaxies is large. The ratio
of S0 galaxies to elliptical galaxies in the cluster is $1.7\pm 0.9$,
which is larger than that in intermediate redshift clusters ($z\sim
0.5$). On the contrary, the fraction of spiral galaxies is only
$24_{-24}^{+27}$\%.  \citet{lub98} estimated that the virial mass of the
cluster is $M_{\rm vir}=7.8_{-2.1}^{+3.2}\times 10^{14} h^{-1}\;\rm
M_{\sun}$, which seems to be exceptionally massive at that redshift. Our
model predicts that at $z\gtrsim 1$, most of the galaxies infalling into
clusters of the mass should be subject to ram-pressure stripping
(Figure~\ref{fig:m_rstvir}). Thus, the high S0 fraction may be due to
the transformation of the field spiral galaxies by ram-pressure
stripping.  If this is the case, the transformation by stripping must be
rapid. This is because the infall rate of field galaxies, most of which
are blue spiral galaxies, increases with $z$ \citep{kau95}; the
ram-pressure stripping must convert rather part of the blue spiral
galaxies into S0 galaxies in a short time. On the other hand, most of
the galaxies in a poor cluster CL~$0023+0423$ at $z=0.84$ are normal
spiral galaxies \citep{lub98}. The population suggests that the
ram-pressure stripping is not effective in the cluster. Since this
cluster consists of two small components of $M_{\rm
vir}=1.0_{-0.4}^{+0.5}\times 10^{13} h^{-1}\rm\; M_{\sun}$ and
$2.6_{-0.8}^{+1.6} h^{-1}\times 10^{14}\rm\; M_{\sun}$, it is
qualitatively consistent with our prediction
(Figure~\ref{fig:m_rstvir}).  More observations of galaxy groups with
$M_{\rm vir}\lesssim 3\times 10^{13}\;\rm M_{\sun}$ at $z\sim 1$ may
give us information about the non-gravitational heating of ICM for
$z\gtrsim 1$(see Figure~\ref{fig:m_rstvir}). However, for the
quantitative comparison between the theory and the observations, we may
need to use a so-called semi-analytic model of cluster formation or a
numerical simulation including the effects of the galaxy infall rate,
the variation of galaxy orbits, and the ram-pressure stripping.

\citet{tan00} found that UV-excess red galaxies are abundant in a rich
cluster near the quasar B2~$1335+28$ at $z=1.086$; the fraction of such
galaxies is relatively small at $z\sim 0.2$ \citep{sma98}. \citet{tan00}
estimated the Abell richness of the cluster and found that it is class
one. By comparing the Abell catalogue \citep{abe58} with X-ray
catalogues \citep[e.g.][]{jon99,moh99,ett99,arn99}, it is shown that the
ICM temperature of the class one clusters is $T_{\rm ICM}\sim
3-4$~keV. If $T_{\rm ICM}=T_{\rm vir}$, the temperature corresponds to
$M_{\rm vir}\sim 5-10\times 10^{14}\:\rm M_{\sun}$ in our models. Thus,
our model predicts that most of the galaxies infalling into the cluster
should be subject to ram-pressure stripping
(Figure~\ref{fig:m_rstvir}). Since the color of the UV-excess red
galaxies can be explained by the superposition of very weak star
formation activity on old stellar population, the galaxies may be the
ones suffering from ram-pressure stripping and the very weak star
formation may be ember although alternative interpretations may be
possible. If this is true, the abundance of the galaxies may also
indicate that ram-pressure stripping happens very often at $z\sim 1$,
although more sophisticated models are required for quantitative
arguments as is the case of CL~$1604+4304$ and CL~$0023+0423$. In the
future, it will be useful to investigate the morphology of the UV-excess
red galaxies in high-redshift clusters in order to know whether the
galaxies are the ones subject to ram-pressure stripping; if the galaxies
have red disks, they are probably the ones \citep{fuj99a}.  It is to be
noted that in a rich cluster MS~1054-03 at $z=0.83$, most of the spiral
galaxies are red \citep{van00}.

At intermediate redshift $0.2\lesssim z \lesssim 0.5$, observations show
that the fraction of blue galaxies in rich clusters, $f_b$, is larger
than that at $z\sim 0$ \citep{but78,but84}. On the other hand, our model
predicts that the fraction of galaxies affected by ram-pressure
stripping increases with $z$. It seems that these contradict to each
other, because the ram-pressure stripping suppresses star formation in
galaxies \citep{fuj99a}. However, as is mentioned above, the infall rate
of field spiral galaxies is higher at the intermediate redshift in
comparison with $z\sim 0$. Thus, the increase of $f_b$ that occurs in
galaxy clusters at the redshift may suggest that the effect of the
increase of the infall rate overwhelms the effect of the ram-pressure
stripping in snapshots of galaxy population, and that the increase of
$f_b$ is not inconsistent with our prediction; we predict that the for a
given mass, ram-pressure stripping at the intermediate redshift is not
as effective as that at $z\sim 1$ (Figure~\ref{fig:m_rstvir}). Moreover,
the average mass of the clusters observed at the intermediate redshift
is expect to be smaller than that at higher redshift ($z\sim 1$),
because the clusters at the intermediate redshift are nearer to us and
easier to be observed. On the other hand, observations also suggest that
the star formation of the infalling galaxies are ultimately truncated in
clusters at the intermediate redshift, although there is a contradiction
over the way the star formation is suppressed
\citep{pog99,bal99b}. Considering that most of the clusters observed at
the intermediate redshift are fairly rich, ram-pressure stripping should
be fairly effective in the clusters, although it is not as effective as
that at $z\sim 1$ (Figure~\ref{fig:m_rstvir}). Thus, our model appears
to be consistent with the truncation of star formation. Moreover, it is
qualitatively consistent with the observed transformation of spiral
galaxies into S0 galaxies in rich clusters at the intermediate redshift
(see \S\ref{sec:intro}).

Except for extremely massive clusters ($M_{\rm vir}\gtrsim 3\times
10^{15}\;\rm M_{\sun}$), Figure~\ref{fig:m_rst} shows that $r_{\rm
st}\lesssim 1$~Mpc regardless of redshift $z$ and the ICM heating. For
nearby clusters at $z\sim 0$, observations show that the fraction of S0
galaxies increases at $r\lesssim 2$~Mpc \citep{whi93,dre97}, which
appears to be larger than the prediction by our ram-pressure model if
the transformation into S0 galaxies is due to ram-pressure
stripping. Moreover, for clusters observed at $0.2\lesssim z\lesssim
0.5$, the fraction of blue galaxies also decreases at $r\lesssim 2$~Mpc
\citep{abr96, van98, mor98, kod00}. If ram-pressure stripping is the
main mechanism of the transformation of blue spiral galaxies into S0
galaxies, the observations suggest that some galaxies at $r\gtrsim
1$~Mpc have already been affected by the stripping when they were in the
progenitors of the clusters before accreted by the main cluster
progenitors. In fact, Figure~\ref{fig:r_stvir} shows that the
ram-pressure stripping in progenitors is effective at least for
$z\lesssim 3$ if the ICM is not heated non-gravitationally and it is
effective for $z\lesssim 1-2$ even if the ICM is heated
non-gravitationally. Moreover, even if galaxies were inside the
stripping radius of the main progenitor at some earlier time, some of
them may have been scattered to large apocenter orbits during the merger
process of cluster progenitors \citep{fuj99b,bal00}.

Analyzing observational data of galaxies in rich clusters, \citet{kod00}
investigated the star formation history of all the galaxies in the
central regions of the clusters. They found that star the formation rate
per galaxy mass declines more rapidly than in the field environment at
$z\lesssim 1$; it suggests the truncation of star formation in most of
the galaxies. This may imply that ram-pressure stripping has been
effective in the clusters or in their progenitors at least for
$z\lesssim 1$, because the star formation rate of galaxies should
decline after the ISM, from which stars are born, is stripped. Since
Figure~\ref{fig:r_stvir} shows that the ram-pressure stripping has been
effective for $z\lesssim 1$ for rich clusters regardless of the
non-gravitational heating, the results of \citet{kod00} are consistent
with our predictions. Unfortunately, because of large uncertainty, their
results cannot constrain the star formation rate of the galaxies in the
cluster progenitors for $z\gtrsim 1$. Thus, we cannot discuss the effect
of the non-gravitational heating (Figure~\ref{fig:r_stvir}).

\section{Conclusions}
\label{sec:conc}

We have studied ram-pressure stripping of galaxies in clusters and their
progenitors. In particular, we pay attention to its dependence on
redshift and the mass of clusters. As a model galaxy, we consider a
radially infalling disk galaxy; the initial position and velocity are
given by a spherical collapse model of structure formation. Since X-ray
observations show that the ICM of nearby clusters is heated
non-gravitationally, we also investigate the effect of the heating on
the ram-pressure stripping.  Our main findings are the following:

1. For a given redshift, ram-pressure stripping of galaxies is more
   effective in more massive clusters. This is because the velocity of
   the radially infalling galaxy increases with the virial mass of the
   cluster. If ram-pressure stripping transforms spiral galaxies into S0
   galaxies, our model is consistent with the observed relation between
   galaxy populations and cluster luminosities (or temperatures).

2. For a given mass of clusters, ram-pressure stripping of galaxies in
   the clusters is more effective at higher redshift. This is because
   the density of the intracluster medium increases with the
   redshift. In particular, at $z\gtrsim 1$, most of the galaxies
   radially infalling into the centers of rich clusters are affected by
   ram-pressure stripping. The relatively high fraction of S0 galaxies
   and the abundance of UV-excess galaxies in rich clusters at $z\sim 1$
   may be due to the ram-pressure stripping.

3. The non-gravitational heating reduces the effect of ram-pressure
   stripping for clusters with $M_{\rm vir}\lesssim 5\times 10^{13}\:\rm
   M_{\sun}$ at $z\sim 0$ and for those with $M_{\rm vir}\lesssim
   10^{13}\:\rm M_{\sun}$ at $z\sim 1$. However, for clusters with
   $M_{\rm vir}\gtrsim 10^{14}\:\rm M_{\sun}$, it does not have an
   influence on the effect of ram-pressure stripping.

4. If the ICM (or the gas accreted by a cluster later on) is heated
   non-gravitationally at $z>>1$, ram-pressure stripping does not occur
   in cluster progenitors at $z\gtrsim 1-2$, because the heat makes the
   ICM fraction of the cluster progenitors small. On the other hand, if
   the ICM had not been heated non-gravitationally until $z\sim 0$,
   ram-pressure stripping occurs even at $z\sim 3$.

\acknowledgments

I am grateful to T. Yamada, M. Nagashima, I. Tanaka, T. Kodama,
T. Tsuchiya, and D. A. Dale for useful discussions and comments.
Comments from an anonymous referee led to significant improvements in
the quality of this paper.

\clearpage

\end{document}